%% file: main.tex
\documentclass[9pt,twocolumn]{article}
\usepackage[left=.75in,right=.75in]{geometry}
\usepackage{booktabs}
\usepackage{dsfont}
\usepackage{amssymb}

\usepackage{multirow}
\usepackage{tikz}
\usepackage{amsmath}
\usepackage{xurl}
\usepackage{threeparttable}
\usepackage{multirow}
\usepackage{placeins}
\usepackage{subcaption}

\usepackage{xcolor}
\usepackage[linesnumbered,ruled,vlined]{algorithm2e}
\usepackage[all]{xy}
\usepackage{hyphenat}
\usepackage{tablefootnote}
\usepackage{titling}
\usepackage{balance}

\usepackage[
	n,
	operators,
	advantage,
	sets,
	adversary,
	landau,
	probability,
	notions,	
	logic,
	ff,
	mm,
	primitives,
	events,
	complexity,
	asymptotics,
	keys]{cryptocode}

\pretitle{\centering\Large\bfseries}
\posttitle{\par\vspace{1em}}

\preauthor{\centering\normalsize\lineskip 0.5em}
\postauthor{\par}

\predate{\centering\small}
\postdate{\par\vspace{1.5em}}

\begin{document}

\title{DeSIA: Attribute Inference Attacks Against Limited Fixed Aggregate Statistics}

\date{}

\author{
Yifeng Mao\footnotemark[1] \quad 
Bozhidar Stevanoski\footnotemark[1] \quad 
Yves-Alexandre de Montjoye
\\[1ex]
\textit{Imperial College London}
}

\maketitle

\def\thefootnote{*}\footnotetext{Equal contribution}\def\thefootnote{\arabic{footnote}}

\begin{abstract}
  \input{sections/0abstract}
\end{abstract}

\input{sections/1introduction}
\input{sections/2background}

\input{sections/3methodology}
\input{sections/4experimental_setup}

\input{sections/5results}
\input{sections/6MIA}
\input{sections/7discussion}

\input{sections/8related_work}
\input{sections/9conclusion}
\input{sections/acknowledgements}

\balance
\bibliography{sections/reference}
\bibliographystyle{plain}

\end{document}

%% file: sections/0abstract.tex
Empirical inference attacks are a popular approach for evaluating the privacy risk of data release mechanisms in practice. While an active attack literature exists to evaluate machine learning models or synthetic data release, we currently lack comparable methods for fixed aggregate statistics, in particular when only a limited number of statistics are released. We here propose an inference attack framework against fixed aggregate statistics and an attribute inference attack called DeSIA. We instantiate DeSIA against the U.S. Census PPMF dataset and show it to strongly outperform reconstruction-based attacks. In particular, we show DeSIA to be highly effective at identifying vulnerable users, achieving a true positive rate of 0.14 at a false positive rate of $10^{-3}$. We then show DeSIA to perform well against users whose attributes cannot be verified and when varying the number of aggregate statistics and level of noise addition. We also perform an extensive ablation study of DeSIA and show how DeSIA can be successfully adapted to the membership inference task. Overall, our results show that aggregation alone is not sufficient to protect privacy, even when a relatively small number of aggregates are being released, and emphasize the need for formal privacy mechanisms and testing before aggregate statistics are released.

%% file: sections/1introduction.tex
\section{Introduction}

Empirical inference attacks are a popular tool to evaluate the privacy risks of data release mechanisms such as machine learning models and synthetic data. Membership inference and attribute inference attacks have been used extensively to evaluate the privacy of machine learning models and synthetic data in practice, showing models to memorize training data~\cite{ponomareva2023dp} and detecting privacy violations~\cite{annamalai2024you}. Data protection laws, such as GDPR~\cite{gdpr}, furthermore define anonymous data as the absence of empirical attacks.

\begin{figure}
    \centering
    \includegraphics[width=1\linewidth]{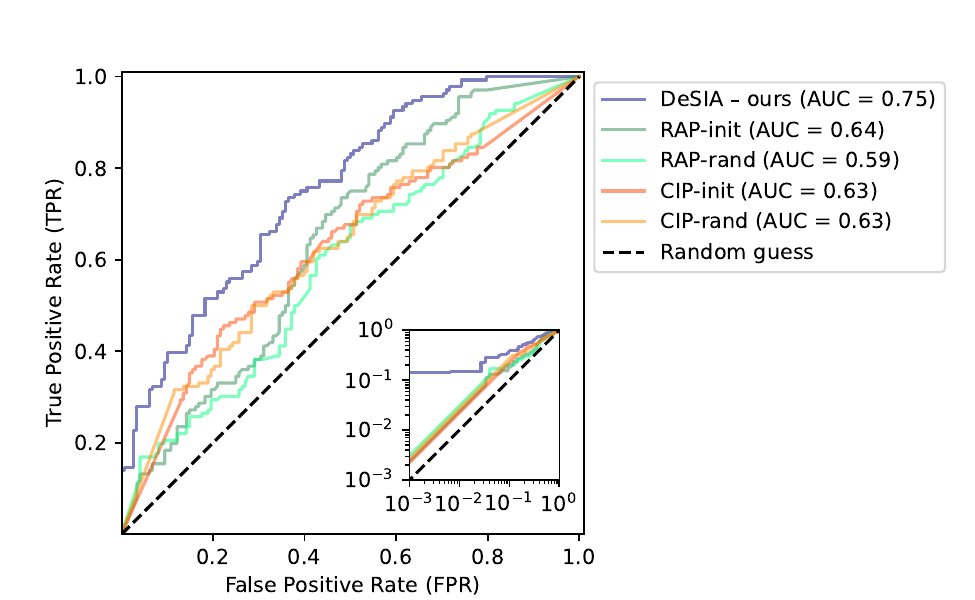}
    \caption{Performance on the PPMF dataset of our attack and the baseline reconstruction-based attacks at predicting the value of the sensitive attribute for all target users. The inset plot shares the same axes as the main plot.}
    \label{fig:tpr_fpr}
\end{figure}

We, however, currently lack a comparable inference attack framework and methods against fixed aggregate statistics, in particular when only a limited number of statistics are released. 
The literature on attacks against fixed aggregate statistics has instead focused on reconstruction attacks.
As of today, the state-of-the-art reconstruction attacks against fixed aggregate statistics are CIP and RAP. CIP, proposed by the the U.S. Census Bureau~\cite{abowd20232010}, models the attack as a constraint integer programming problem. RAP, proposed by Dick et al.~\cite{dick2023confidence}, uses synthetic data to reconstruct tabular data from aggregate statistics. 

\textbf{Contributions}. We here propose a framework and a method for attribute inference attacks against fixed aggregate statistics from tabular data. Our method, DeSIA, consists of two modules: a deterministic module, which we instantiate as a modified constraint integer programming problem, and a stochastic module, which we use to infer the most likely attribute when the deterministic module cannot verify its solution.

We instantiate our attack against the PPMF release from the U.S. Census Bureau and show DeSIA to strongly outperform state-of-the-art reconstruction-based attacks when instantiated on the attribute inference task. We also show DeSIA to successfully identify highly vulnerable users achieving a $0.14$ true positive rate (TPR) at a false positive rate (FPR) of $10^{-3}$. We then show DeSIA to perform well a) against users whose attributes cannot be verified by the deterministic module, b) when varying the number of aggregate statistics, and c) when varying the level of per-query noise addition. We also perform an extensive ablation study of DeSIA, showing all the components to be necessary for it to perform well. We finally show how DeSIA can be successfully adapted to the membership inference task (MIA) here again strongly outperforming reconstruction-based baselines. Overall, our results show practical information leakage to occur even when a relatively small number of aggregates are being released, and emphasize the need for formal privacy mechanisms and testing before aggregate statistics are released.

%% file: sections/2background.tex
\section{Background}

\subsection{Fixed Aggregate Statistics}
We consider a data curator handling a private tabular dataset $D$ from a distribution $\mathcal{D}$. The dataset $D$ consists of records for a set of $s$ users $U$, $s=|U|$, over $n$ attributes, $A=\{a_1, \dots, a_n\}$. Each attribute $a_i \in A$ can take values from a set $\mathcal{V}_i, i\in\{1,\dots,n\}$. We denote the record of user $u \in U$ as $r_u=(r_u^1, \dots, r_u^n)$, where $r_u^i$ is a value from the set $\mathcal{V}_i$, $r_u^i \in \mathcal{V}_i, i\in\{1,\dots,n\}$. We denote the projection of the user record $r_u$ over a subset of attributes $A'=\{a_{i_1}, \ldots, a_{i_k}\} \subseteq A$ as $r_u^{A'} = (r_u^{i_1}, \dots, r_u^{i_k})$.

The data curator aims to release aggregate statistics from the dataset $D$, where each aggregate statistic counts the number of records in particular subsets of $D$. Formally, we define a counting aggregate statistic $q$ as $q = (V_1^q, \dots, V_n^q)$, given sets $V_i^q$ of values corresponding to each attribute $a_i$, $V_i^q \subseteq \mathcal{V}_i$. We denote the subset of $U$ as $U_q$=$\{u | r_u^1 \in V_1^q \land \dots \land r_u^n \in V_n^q, \forall u \in U\}$ and we call $U_q$ the userset of the aggregate statistic $q$. We denote the result of $q$ evaluated on $D$ by $q(D)$, $q(D) = |U_q| = \sum_{u\in U}\mathds{1}(r_u^1 \in V_1^q \land \dots \land r_u^n \in V_n^q)$. Equivalently, the aggregate statistic $q$ can be expressed as a counting query using SQL notation as:
\begin{gather}\label{eq:general_syntax}
\begin{aligned}
    \text{SELECT} & \; \text{count(*)} \;\text{FROM} \; D \\
    \text{WHERE} & \; a_1 \; IN \; V_1^1 \; AND \; \ldots \;   AND \; a_{n-1} \; IN \; V_{n-1}^1 \\ \; & AND \; a_n \; IN \; V_n^1.
\end{aligned}
\end{gather}

\subsection{Threat Model: Attribute Inference Attacks Against Fixed Aggregate Statistics}
We evaluate the privacy leakage of releasing fixed aggregate statistics from a tabular dataset in the context of attribute inference attacks (AIAs). 

We assume the data curator releases a fixed set of $m$ aggregate statistics, $Q=\{q_1, \dots, q_m\}$ and their values on the dataset $D$, $Q(D) = \{q_1(D),$ $\dots,$ $q_m(D)\}$.
We assume that fewer aggregates are released than the number of users in $D$, $m < s$, i.e., that on average there is fewer than one aggregate per user, $\frac{m}{s} < 1$. We evaluate various values of $\frac{m}{s}$ in Section~\ref{sec:num_queries}.

We assume that one of the attributes is a sensitive attribute. Without loss of generality, we assume $a_n$ is sensitive and denote the set of non-sensitive attributes as $A' = \{a_1, \ldots, a_{n-1}\}$.

\textbf{Attacker's Goal}. The attacker's goal is to infer the value of the sensitive attribute $a_n$ of a target user $u^*$, i.e., to infer $r_{u^*}^n$.

\textbf{Attacker's knowledge}. In line with the literature, we assume the attacker to have auxiliary knowledge about the dataset $D$ and knowledge about the target user ${u^*}$.

The auxiliary knowledge of the dataset $D$ consists of knowledge of its size $s$ and access to an auxiliary dataset $D_{aux}$ from a distribution $\mathcal{D}_{aux}$, similar to $\mathcal{D}$. We also assume the attacker to know the possible values for every attribute $a_i \in A, i\in\{1,\dots,n\}$.

Knowledge about the target user $u^*$ consists of the values of the non-sensitive attributes $A'$ of the target user $u^*$, $r_{u^*}^{A'} = (r_{u^*}^1, \dots, r_{u^*}^{n-1})$ and the knowledge that the target user is unique in the dataset $D$ with those values for the non-sensitive attributes, $\forall u' \in U, u' \neq u^*, r_{u'}^{A'} \neq r_{u^*}^{A'}$.

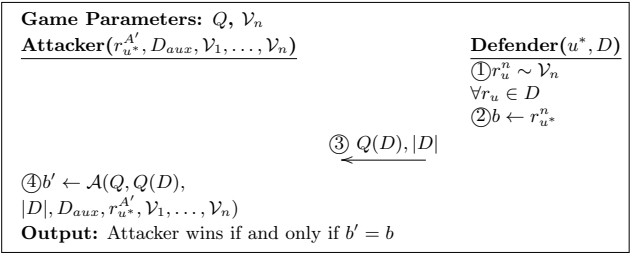
\begin{figure}
\resizebox{.475\textwidth}{!}{
\fbox{\small
\begin{minipage}{1.2\columnwidth}
\begin{tabular}{lcl}
\textbf{Game Parameters: $Q$, $\mathcal{V}_n$} & & \\
\textbf{\underline{Attacker($r_{u^*}^{A'}, D_{aux}, \mathcal{V}_1, \dots, \mathcal{V}_n$)}} & & \textbf{\underline{Defender($u^*, D$)}}\\
 & & $\textcircled{1} r_u^n \sim \mathcal{V}_n$ \\
 & & $\forall r_u \in D$  \\
 & & $\textcircled{2} b \leftarrow r_{u^*}^n$ \\
 & $\xymatrix@1@=40pt{& \ar[l]_*{\textcircled{3} \ Q(D), |D|}}$ & \\
 $\textcircled{4} b' \leftarrow \mathcal{A}(Q, Q(D),$ & & \\
 $|D|, D_{aux}, r_{u^*}^{A'}, \mathcal{V}_1, \ldots, \mathcal{V}_n)$ & & \\
\end{tabular}
\begin{tabular}{l}
\textbf{Output:} Attacker wins if and only if $b' = b$
\end{tabular}
\end{minipage}
}}
\caption{Privacy game for attribute inference attack against fixed aggregate statistics.}
\label{fig:privacy_game_aia}
\end{figure}

\subsection{Framework for Attribute Inference Attack as a Privacy Game} 
We propose a formal framework for attribute inference attacks against fixed aggregates as a privacy game. The privacy game (Fig.~\ref{fig:privacy_game_aia}) is played between two players, an attacker and a defender, over a total of four steps. The defender plays the first three steps and the attacker plays the fourth one.

First, in line with the literature on attacks against query-based systems~\cite{cretu2022querysnout,stevanoski2024querycheetah}, the defender samples the values of the sensitive attribute uniformly at random from $\mathcal{V}_n$ for all users $U$ in the protected dataset $D$. This provides a random baseline and circumvents the imputation issue (see Section~\ref{sec:related_work}), allowing us to measure the privacy risk posed solely by the release of the aggregate statistics.

Second, the defender saves the value that was sampled for the sensitive attribute of target user $u^*$, which we denote as $b$. Note that the value $b$ is unambiguous as we assume, for simplicity and in line with the literature, that the target user $u^*$ is unique in $D$ for $r_{u^*}^{A'}$.

Third, the defender evaluates the aggregate statistics $Q$ on the protected dataset $D$ with randomized values of the sensitive attribute and releases them to the attacker. 

Fourth, the attacker aims to predict the sensitive value of the target user $b$ by using the released aggregate statistics. The attacker runs an attack $\mathcal{A}$ to obtain a value $b'$.

The attacker wins the game if and only if the predicted value $b'$ is equal to the value $b$ of the target user's sensitive attribute in $D$, $b'=b$.

\subsection{State-of-the-art: Reconstruction-Based Attacks against Fixed Aggregate Statistics}
Existing literature on empirical attacks against fixed aggregate statistics has primarily focused on reconstruction attacks. Reconstruction attacks aim to reconstruct the private dataset $D$ by using the released statistics $Q$ and their values on $D$, $Q(D)$. They search for a dataset $D'$ such that the values of the released aggregate statistics $Q$ from $D'$, $Q(D')$, exactly or closely match the released values from the protected dataset $D$, $Q(D') \approx Q(D)$, with the ideal goal being perfect equality $Q(D') = Q(D)$. The dataset $D'$ is called a tentative reconstructed dataset.

The state-of-the-art attacks are an attack that uses constraint integer programming~\cite{abowd20232010} (CIP) and an attack using synthetic data generation techniques~\cite{dick2023confidence} to search for a tentative reconstructed dataset (RAP).

\subsubsection{CIP: Constraint Integer Programming Attack} 

Abowd et al.~\cite{abowd20232010} proposed an attack, which we refer to as CIP, that aims to find a tentative reconstructed dataset $D'$ by formulating a constraint integer programming task. The main idea of CIP is to represent a dataset $D'$ by using the multiplicity of all potential records $(v_1, \dots, v_n), \ \forall v_i \in \mathcal{V}_i$ in $D'$ as integer variables $x_{(v_1, \dots, v_n)}$. For example, a record that is not a member of $D'$ has a multiplicity of $0$. Note that while Abowd et al.~\cite{abowd20232010} have modified this problem formulation to an equivalent binary variable formulation for a more efficient execution time, for simplicity, we here present and use the representation with integer variables.

Formally, the constraint integer programming task is composed of three main components: a set of unknown variables $X$, sets of possible values for each unknown variable (also called domain sets) $S$, and constraints $C$ that specify allowable combinations of unknown variables.

Each \textbf{unknown variable}, $x_{(v_1, \dots, v_n)} \in X$, represents the multiplicity of a record $(v_1, \dots, v_n), v_i \in \mathcal{V}_i$ in $D'$. Note that although there are $\prod_{i=1}^n |\mathcal{V}_i|$ unknown variables, many of them may have values of $0$ since they can represent records that are not members of the dataset $D'$.

Each \textbf{domain}, $s_{(v_1, \dots, v_n)} \in S$, contains all possible values for the unknown variable $x_{(v_1, \dots, v_n)}$. Here, a domain contains all possible integers between $0$ and the size of the protected dataset $D$, inclusively, $x_{(v_1, \dots, v_n)} \in \{0,1,\dots, s\}$

Each \textbf{constraint} $c_q \in C$, represents an allowable combination of variables given a released aggregate statistic $q = (V_1^q, \dots, V_n^q),$ $q\in Q$. The constraint $c_q$ limits the possible values for the multiplicity of records by taking the aggregate value $q(D)$: $$c_q: \sum\limits_{\substack{\forall r \in \mathcal{V}_1^q \times \cdots \times \mathcal{V}_n^q}} x_r = q(D).$$

The solution to the constraint integer programming task is an assignment function that assigns values to all unknown variables. We denote the assignment function as $\sigma: X \rightarrow \mathbb{Z}$. Abowd et al. use a third-party solver~\cite{gurobi} to obtain a tentative reconstructed dataset $D'$, which we also use.

We consider two variants of this attack, depending on the initialization of the third-party solver. The attack originally proposed by Abowd et al. \cite{abowd20232010}, which we refer to as \textbf{CIP-rand}, where the third-party solver initializes all unknown variables to zeros by default. For a fair comparison, we also use an adaptation of the attack, which we refer to as \textbf{CIP-init}, that uses the access of the auxiliary dataset $D_{aux}$ to initialize the unknown variables $x_{(v_1, \dots, v_n)}$ in the third-party solver with the multiplicity of the records in the auxiliary dataset $D_{aux}$.

\subsubsection{RAP: Synthetic Data Attack} 
Dick et al.~\cite{dick2023confidence} proposed an attack, which we refer to as RAP, that produces a tentative reconstructed dataset $D'$ by using a synthetic data generator model. 

The attack trains a synthetic data generator model $M$ that takes as input aggregate statistics $Q$ and their values $Q(D)$ on the protected dataset $D$ and outputs a tentative reconstructed dataset $D'$, The model $M$ is trained to minimize the L2-distance between the aggregate values on the synthetic dataset $D'$ and the protected dataset $D$, $\text{min}\lVert Q(D') - Q(D) \rVert_2$.

We use the two variants of this attack proposed by Dick et al., depending on the initialization of model $M$: \textbf{RAP-rand} where $M$ is initialized with a uniformly at random sample from $\mathcal{V}_1 \times \dots \times \mathcal{V}_n$, and \textbf{RAP-init} where $M$ is initialized with the auxiliary dataset $D_{aux}$.

\begin{figure*}[t]
    \centering
    \includegraphics[width=0.75\linewidth]{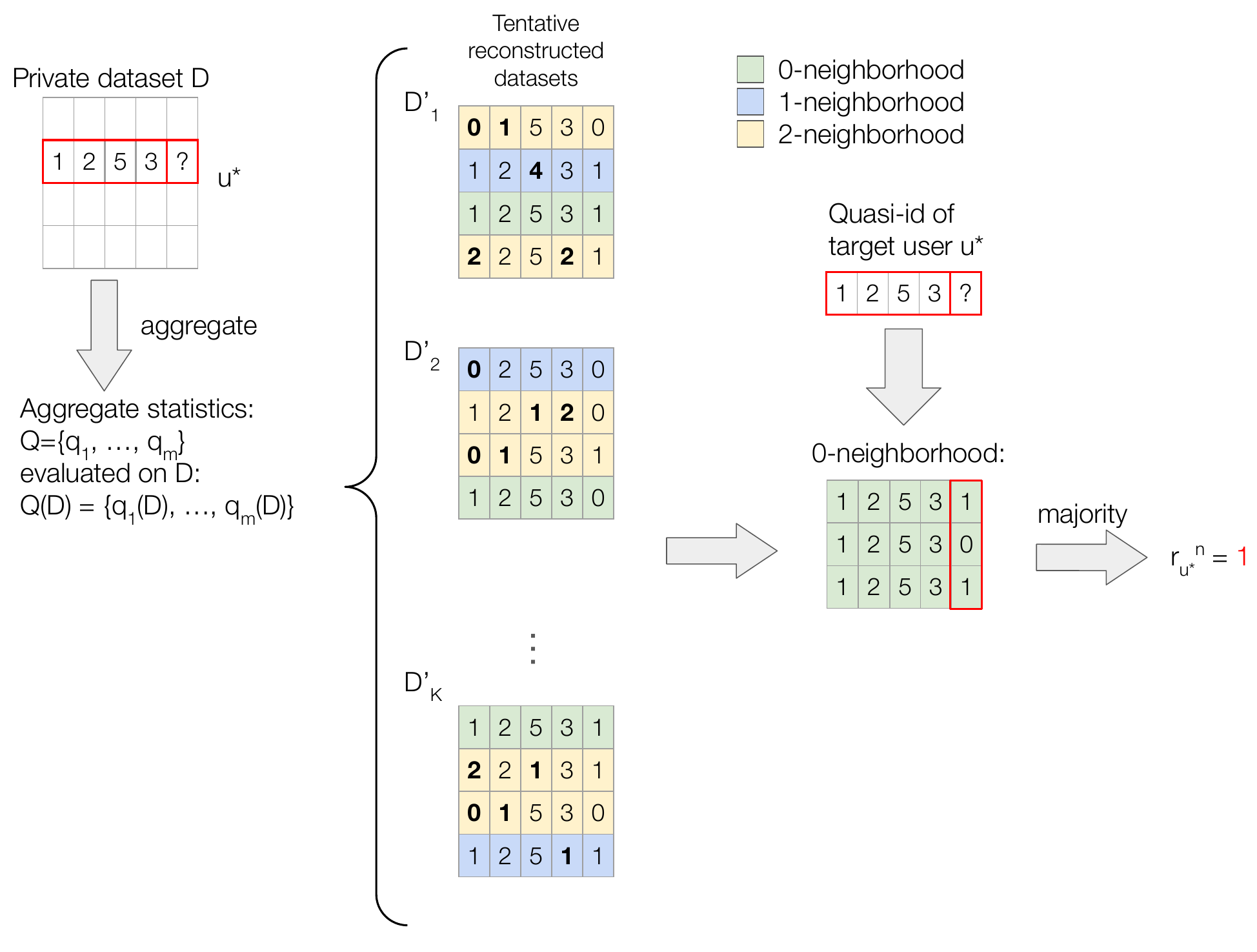}
    \caption{Adaptation of reconstruction attacks for attribute inference.}
    \label{fig:adaptation}
\end{figure*}

We would like to note first that while a perfect reconstruction attack would fully reconstruct the private dataset $D$ and thus be a perfect attribute inference attack, this is rarely the case in practice, including when a large number of aggregate statistics are released, e.g., by the U.S. Census Bureau. Second, we assume, in line with the broader literature on inference attack, an attacker weaker than the strong Differential Privacy (DP) attacker but with access to an auxiliary dataset. For fair comparison, we adapt the CIP attack to take advantage of the auxiliary dataset. The RAP attack already proposes a variant leveraging an auxiliary dataset which we use here. Third, we assume an attacker with access to at least some information about the target record, but note that, while this is necessary for any attribute or membership inference attack, some threat models using reconstruction attacks do not need this assumption, e.g. identifying that an individual with a certain characteristic lives in a block. We do not consider this threat here.

\subsection{Instantiating Reconstruction Attacks for Attribute Inference}\label{sec:ra_to_aia}
In most cases, in particular when limited number of aggregate statistics are released, there are more than one tentative reconstructed datasets $D'$~\cite{abowd20232010}. We adapt existing reconstruction attack to the attribute inference task by (a) generating $K$ tentative reconstructed datasets $D'_1, \dots, D'_K$ and (b) taking a majority vote after combining the values of the target user's sensitive attribute from all $K$ datasets. Figure~\ref{fig:adaptation} illustrates both steps.

\textbf{Obtaining tentative reconstruction datasets}. For RAP, we follow the approach by Dick et al.~\cite{dick2023confidence} and train $K$ synthetic data generator models, each with a different random seed, and generate a tentative reconstructed dataset using each of the $K$ models. For CIP, we follow a similar approach by applying the solver $K$ times, each time with a different seed and order of constraints. 

\textbf{Majority vote in smallest existing neighborhood}. We define $L-neighborhood$ as the records from all $K$ tentative reconstructed datasets that differ at $L$ values from the values of the non-sensitive attributes of the target user, $r_{u^*}^{A'}$:
\begin{equation*}
    \begin{split}
        L-&neighborhood(r_{u^*}^{A'}, D'_1,\dots,D'_K) = \\
        &= \{r | \sum_{i=1}^{n-1}\mathds{1}(r^i \neq r_{u^*}^i) = L, r \in \bigcup_{j\in\{1,\dots,K\}}D'_j\}.
    \end{split}
\end{equation*}
For example, the 0-neighborhood is the set of records from all $K$ tentative reconstructed datasets that have the same values for the non-sensitive attributes $A'$ as the target user, $0-neighborhood = \{r| r^{A'} = r_{u^*}^{A'} r \in \bigcup_{j\in\{1,\dots,K\}}D'_j \}$.

For both RAP and CIP, we start from the 0-neighborhood of the target user. The attacker infers the sensitive attribute of the target user, $r_{u^*}^n$, by predicting the most frequent value for attribute $a_n$ of all records in the 0-neighborhood.
$$r_{u^*}^n =   \arg\max count(r^n).$$
$$s.t. \ r \in 0-neighborhood(r_{u^*}).$$ 
If the $0-neighborhood$ is empty, we extend to the $1-neighborhood$. Following the same principle, if the $(L-1)-neighborhood$ is empty, we extend to $L-neighborhood$, until $L = n-1$, when we take all records in all tentative reconstructed datasets $\bigcup_{i\in\{1,\dots,K\}}D'_i$.

If there is a tie for the most frequent value between two (or more) values in a given $L-neighborhood$, we select one uniformly at random from them.

%% file: sections/3methodology.tex
\section{Methodology}

\begin{figure*}
    \centering
    \includegraphics[width=0.8\linewidth]{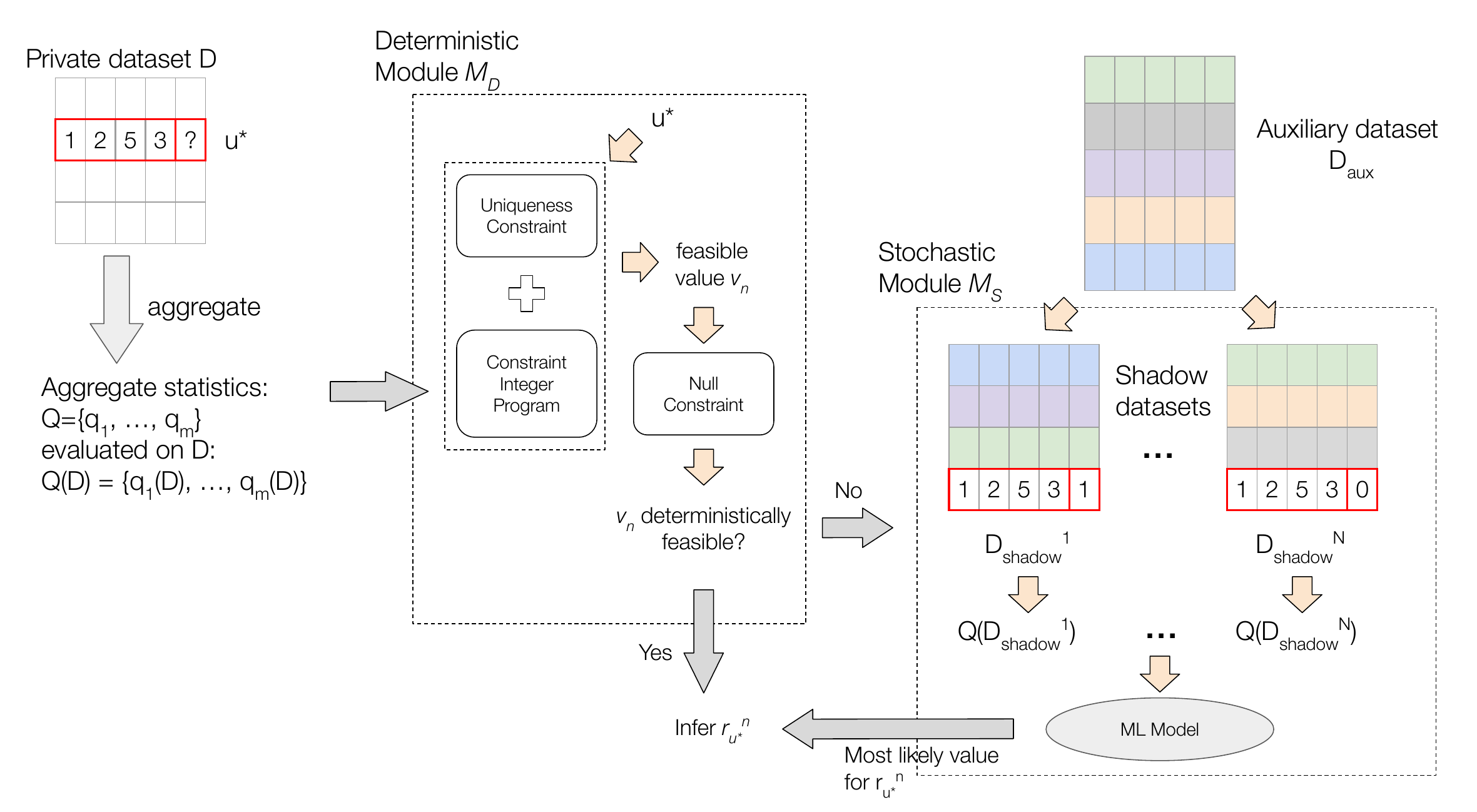}
    \caption{Our proposed attack. It uses two modules: (1) $\mathcal{M}_D$ deterministically checking if there is only one possible value for the target user's sensitive attribute, and (2) $\mathcal{M}_S$ stochastically predicting the most likely value.}
    \label{fig:methodology}
\end{figure*}

Our method, DeSIA (Deterministic-Stochastic Inference Attack) is composed of two modules: a deterministic module $\mathcal{M}_D$ and a stochastic module $\mathcal{M}_S$. The deterministic module $\mathcal{M}_D$, which we instantiate as a modified constraint integer programming problem, aims to identify if one and only one value for the sensitive attribute of the target user is deterministically feasible, given the released statistics, and to output that value. If there is more than one feasible value, the stochastic module $\mathcal{M}_S$ is then used to estimate the likelihood of the feasible values and to output the most likely one.

 \[ \text{DeSIA} = \begin{cases} 
          \mathcal{M}_D, & \mathcal{M}_D(\cdot) \neq \varnothing\\
          \mathcal{M}_S, & otherwise
       \end{cases}\]

\subsection{Deterministic Attribute Inference}

The deterministic module $\mathcal{M}_D$ first finds a feasible value $v^*_n\in\mathcal{V}_n$ for the sensitive attribute of the target user, $r_{u^*}^n$, and second, deterministically verifies that the value $v^*_n$ is the only feasible value. We instantiate both tasks as constraint integer programming tasks (see Algorithm~\ref{algo:deterministic}).

\textbf{Finding a feasible value}. We use constraint integer programming to obtain a possible value $v^*_n\in\mathcal{V}_n$ for the target user $u^*$. In particular, we define the unknown variables $x_r$ for record $r = (v_1, \dots, v_n), v_i\in\mathcal{V}_i,i\in\{1,\ldots,n\}$ following the CIP attack. We update the domain of each unknown variable to optimize the execution time and we extend the set of constraints.

While we use the simplified integer representation, we optimize the execution time of the method by updating the domain of each unknown variable to lower the range of possible values, using information from the released aggregate statistics. We set the maximal integer $s_r^{max}$ in the domain set of the unknown variable $x_r$ to be the minimal value of all relevant aggregate statistics:
$$s_r^{max} = min(s, \min\limits_{\substack{q \in Q \\ v_i\in V_i^q, \forall i\in\{1,\dots,n\}}} q(D)).$$

\begin{algorithm}
\small
\caption{\textsc{Deterministic attribute inference}}
\label{algo:deterministic}

\SetCommentSty{mycommfont}
\SetAlgoNoLine
\SetKwInput{KwInput}{Input}                
\SetKwInput{KwOutput}{Output}              
\DontPrintSemicolon
  \KwInput{Aggregate statistics $Q=\{q_1, \ldots, q_m\}$ \;
  \hphantom{Input:  } Aggregate values $Q(D)=\{q_1(D), \ldots, q_m(D)\}$ \;
  \hphantom{Input:  } Dataset size $s$ \;
  \hphantom{Input:  } Attribute domains $\{\mathcal{V}_1, \ldots, \mathcal{V}_n\}$ \;
  \hphantom{Input:  } Non-sensitive attributes $(r_{u^*}^1, \ldots, r_{u^*}^{n-1})$ \;
  }
  \KwOutput{Prediction on $r_{u^*}^n$ if deterministically possible, otherwise NaN.}
Unknown Variable Set $X \leftarrow \emptyset$, Domain Set $S \leftarrow \emptyset$, Constraint Set $C \leftarrow \emptyset$ \tcp*{Initialize}

\For{$r \in \mathcal{V}_1 \times \mathcal{V}_2 \times \cdots \times \mathcal{V}_n$} {
  \textbf{Define} $x_r$ \tcp*{Define the new variable}
  $s_r^{max} \leftarrow \min(s, \min\limits_{\substack{\forall q_i \in Q \\ r \in \mathcal{V}_1^{q_i} \times \cdots \times \mathcal{V}_n^{q_i}}} q_i(D))$ \tcp*{Get minimal upper bound for the new variable}
  $X \leftarrow X \cup \{x_r\}, S \leftarrow S \cup \{0, \ldots, s_r^{max}\}$ \tcp*{Update the variable and domain set}
}

\For{$q_i \in Q$ and $q_i(D)$} {
  $c_i : \sum\limits_{r \in \mathcal{V}_1^{q_i} \times \cdots \times \mathcal{V}_n^{q_i}} x_r$ = $q_i(D)$ \tcp*{Define constraint of $q_i$}
  $C \leftarrow C \cup \{c_i\}$ \tcp*{Update the constraint set}
}

$c_t : \sum\limits_{v_n \in \mathcal{V}_n} x_{(r_{u^*}^1, \dots, r_{u^*}^{n-1}, v_n)}$ = 1\tcp*{Define uniqueness constraint}
$C \leftarrow C \cup \{c_t\}$ \; 

$\sigma \leftarrow \textbf{RunSolver}(X, S, C)$ \;
\If{$\exists \ \sigma$ \tcp*{If there is at least one feasible solution}} {
  \For{$v_n \in \mathcal{V}_n$} {
    \If{$\sigma(x_{(r_{u^*}^1, \dots, r_{u^*}^{n-1}, \hat{v_n})})$ = 1} {
      $c_{t'} : x_{(r_{u^*}^1, \dots, r_{u^*}^{n-1}, \hat{v_n})}$ = 0 \tcp*{Define the null constraint}
      $C' \leftarrow C \cup \{c_{t'}\}$ \tcp*{Update the constraint set}
    }
  }
  $\sigma' \leftarrow \textbf{RunSolver}(X, S, C')$ \tcp*{If there is no feasible solution}
  \If{$\nexists \ \sigma'$} {
    \Return $\hat{v_n}$ \tcp*{Return the only possible value of sensitive attribute}
  }
} 
\Return $\varnothing$
\end{algorithm}

We crucially also extend the set of constraints by adding a constraint specific to the target user $u^*$, which we call a uniqueness constraint. The constraint reflects the attacker's knowledge of (a) the values of the nonsensitive attributes of the target user, $r_{u^*}^{A'}$,  and (b) the uniqueness of the in $D$ given $r_{u^*}^{A'}$:
$$c_t : 1=\sum_{v_n \in \mathcal{V}_n} x_{(r_{u^*}^1, \dots, r_{u^*}^{n-1}, v_n)}.$$

We then use a third-party solver~\cite{gurobi} to solve the constraint integer problem. The solver returns the assignment of values $\sigma$ as a solution, which maps from an unknown variable $x_{r}$ to its assigned value $\sigma(x_{r})$. 

We select the variables $\{x_{(r_{u^*}^1, \dots, r_{u^*}^{n-1}, v_n)} | \ \forall v_n \in \mathcal{V}_n \}$, which correspond to records that share the same values on non-sensitive attributes as the target user. Because of the uniqueness constraint, $\sigma$ only assigns one of the selected variables to be larger than zero. We denote the value of the sensitive attribute of that variable as $\hat{v_n}$. 

\textbf{Verifying the feasible value}. The feasible value $\hat{v_n}$ found for the sensitive attribute of the target user $r_{u^*}^n$ might not be unique. We here formulate another constraint integer task to verify whether the obtained value $\hat{v_n}$ is the only feasible value for $r_{u^*}^n$, explicitly disallowing $\hat{v_n}$ as a possible value for $r_{u^*}^n$ through a null constraint:
$$c_{t'} : 0 = x_{(r_{u^*}^1, \dots, r_{u^*}^{n-1}, \hat{v_n})}.$$

If we do not obtain any feasible assignment function $\sigma'$, $\mathcal{M}_D$ verifies that the value $\hat{v_n}$ is the only deterministically possible value for $r_{u^*}^n$. If this is the case, we consider the target user $u^*$ to be deterministically vulnerable and assign it a certainty of 1. If it is not, we use the stochastic module $\mathcal{M}_S$ to infer an attribute value.

\subsection{Stochastic Attribute Inference}
The stochastic module $\mathcal{M}_S$ estimates the likelihood of the feasible values for $r_{u^*}^n$ and outputs the most likely one (see Algorithm~\ref{algo:stochastic}). In particular, it uses a machine learning model trained on aggregate statistics from datasets similar to the protected dataset $D$. The stochastic module is composed of three steps.

First, we create $N$ datasets, $D_{shadow}^1, \ldots, D_{shadow}^N$, which we call shadow datasets. Each shadow dataset $D_{shadow}^i, i\in\{1,\ldots,n\}$ consists of $s-1$ users sampled without replacement from the auxiliary dataset $D_{aux}$ and the target user $u^*$. We project their records on non-sensitive attributes $A'=\{a_1,\ldots,a_{n-1}\}$ to obtain $\{r_{i, 1}^{A'}, \ldots, r_{i, s-1}^{A'}\}$. Then, we sample $s-1$ values from $\mathcal{V}_n$ uniformly at random, denoted as $\{z_{i,1}, \ldots, z_{i,s-1}\}$. We append $z_{i,j}$ to $r_{i, j}^{A'}$, such that $z_{i,j}$ is the value for the sensitive attribute of the record $(r_{i, j}^{A'}, z_{i,j})$. We obtain a shadow record of the target user by appending a value $z_{i,*}$ sampled at uniform random from $\mathcal{V}_n$ to the values of non-sensitive attributes $r_{u^*}^{A'}$. The resulting shadow dataset $D_{shadow}^i$ is as follow: 
$$\{(r_{i, 1}^{A'},z_{i,1}), \ldots, (r_{i, s-1}^{A'}, z_{i,s-1}), (r_{u^*}^{A'}, z_{i,*})\}.$$ 

\begin{algorithm}
\small
\caption{\textsc{Stochastic attribute inference}}
\label{algo:stochastic}

\SetCommentSty{mycommfont}
\SetAlgoNoLine
\SetKwInput{KwInput}{Input}                
\SetKwInput{KwOutput}{Output}              
\DontPrintSemicolon
  \KwInput{Aggregate statistics $Q=\{q_1, \ldots, q_m\}$ \;
  \hphantom{Input:  } Aggregate values $Q(D)=\{q_1(D), \ldots, q_m(D)\}$ \;
  \hphantom{Input:  } Dataset size $s$ \;
  \hphantom{Input:  } Attribute domains $\{\mathcal{V}_1, \ldots, \mathcal{V}_n\}$ \;
  \hphantom{Input:  } Non-sensitive attributes $(r_{u^*}^1, \ldots, r_{u^*}^{n-1})$ \;
  \hphantom{Input:  } Auxiliary dataset $D_{\text{aux}}$ \;
  \hphantom{Input:  } Number of shadow datasets $N$ \;
  }
  \KwOutput{Prediction on $r_{u^*}^n$, based on the probability of each possible value in $\mathcal{V}_n$}

\For{$i = 1, \ldots, N$} {
  $D^i_{shadow} \leftarrow \emptyset$, $R^i = \emptyset$ \tcp*{Initialize shadow dataset and the set of selected records}
  \For{$j=1, \ldots, s-1$} {
    $r_{i,j} \sim (D_{\text{aux}} \setminus R^i)$ \tcp*{Sample one record without replacement from auxiliary dataset}
    $z_{i,j} \sim \mathcal{V}_n$ \tcp*{Sample the sensitive value at uniform random}
    $R^i \leftarrow R^i \cup \{r_{i,j}\}$ \tcp*{Add the sampled record to selected record set}
    $D^i_{shadow} \leftarrow D^i_{shadow} \cup \{(r_{i,j}^{A'}, z_{i,j})\}\}$ \tcp*{Update shadow dataset}
  }
  $z_{i,*} \sim \mathcal{V}_n$ \;
  $D^i_{shadow} \leftarrow D^i_{shadow} \cup \{(r_{u^*}^{A'}, z_{i,*})\}$ \tcp*{Add shadow record of target user to shadow dataset}
  $Q(D^i_{shadow}) \leftarrow \{q_1(D^i_{shadow}), \ldots, q_m(D^i_{shadow})\}\}$ \tcp*{Evaluate the aggregate statistics on shadow dataset}
}
$M \leftarrow \{(Q(D^1_{shadow}, z_{1,*}),  \ldots, (Q(D^N_{shadow}, z_{N,*})\}$ \tcp*{Train meta-classifier}
$r_{u^*}^{n} \leftarrow M(Q(D))$ \tcp*{Predict the sensitive value of target user using trained meta-classifier}
\Return $r_{u^*}^{n}$
\end{algorithm}

Second, we obtain the fixed aggregate statistics from the $N$ shadow datasets $Q(D_{shadow}^1), Q(D_{shadow}^2), \ldots, Q(D_{shadow}^N)$ by evaluating the released aggregate statistics $Q$ on each shadow dataset. Third, we train a machine learning model $M$ which takes $Q(D_{shadow}^1), $ $Q(D_{shadow}^2), \ldots, Q(D_{shadow}^N)$ as input and the values $z_{1,*}, \ldots, z_{N,*}$ as labels. The model $M$ aims to predict the sensitive attribute of the target user's record. We finally input the aggregates $Q(D)$ of the protected dataset $D$ to the trained machine learning model and use it to infer a predicted value of the sensitive attribute for the target user in $D$, $r_{u^*}^n$.

%% file: sections/4experimental_setup.tex
\section{Experimental setup}

\subsection{Dataset and Aggregates}

\textbf{Privacy-Protected Microdata File (PPMF)}. In line with the literature~\cite{dick2023confidence}, we evaluate our method on the 2020-05-27 vintage Privacy-Protected Microdata File (PPMF)~\cite{uscensusdas}. The PPMF dataset contains record-level data from the U.S. Census Bureau for users across the U.S., partitioned by census block. We use the same subset of the released block-level aggregates in the U.S. Decennial Census as~\cite{dick2023confidence}. In particular, as the U.S. Census Bureau releases the aggregate statistics as contingency tables, we also use the same block-level contingency tables as the literature. We list and briefly describe them in Table~\ref{tab:census_blocks}.

\begin{table}
\centering
\resizebox{.475\textwidth}{!}{
\begin{tabular}{ll}
\toprule
\textbf{Table} & \multirow{2}{*}{\textbf{Description}} \\ 
\textbf{name} &  \\ 
\midrule
P1. & total population \\
P6. & race (total races tallied) \\
P7. & Hispanic/Latino by race (total races tallied) \\
\multirow{1}{*}{P9.} & Hispanic/Latino by race \\
\multirow{1}{*}{P11.} & Hispanic/Latino by race for the population aged 18+\\
P12. & sex by age \\
\midrule
P12A. & sex by age for only race White \\
P12B. & sex by age for only race Black or African American \\
P12C. & sex by age for only race American Indian\\
 &  and Alaska native \\
P12D. & sex by age for only race Asian \\
P12E. & sex by age for only race native Hawaiian \\
 & and other Pacific Islander \\
P12F. & sex by age for only other races \\
P12G. & sex by age for only two or more races \\
P12H. & sex by age for only Hispanic/Latino \\
P12I. & sex by age for only race White and not Hispanic/Latino \\
\bottomrule
\end{tabular}}
\caption{Block-level contingency tables that we use. The same contingency tables were also used by RAP~\cite{dick2023confidence}.}
\label{tab:census_blocks}
\end{table}

\begin{table}[t]
\centering
\begin{threeparttable}
\resizebox{.475\textwidth}{!}{
\begin{tabular}{llr}
\toprule
\textbf{Attribute name} & \textbf{Type} & \textbf{Domain} \\ \midrule
Block ID & Integer & A constant number for each block \\
Gender                       & Boolean & \{Male, Female\} \\
Age                          & Integer & \{0, 1, \ldots, 115\} \\
Race                         & Integer & \{0, 1, \ldots, 63\} \tnote{1} \\ 
Hispanic                     & Boolean & \{Hispanic, Not-Hispanic\} \\
\bottomrule
\end{tabular}}
\begin{tablenotes}
  \footnotesize
  \item[1] Each number represents one race category defined by the U.S. \\ Office of Management and Budget Standards. \cite{OMB1997}
\end{tablenotes}
\caption{Description of attributes in 2010 Census Decennial Microdata.}
\label{tab:census_attributes}
\end{threeparttable}
\end{table}

Table~\ref{tab:census_attributes} presents the attributes, their types, and possible values in the PPMF dataset. We consider the attribute ``Hispanic'' to be the sensitive attribute and the other attributes of the target user to be known by the attacker. Note that we randomize the values of the sensitive attribute so as to avoid the imputation problem and ensure no correlation exists between the sensitive attribute and the other attributes. This allows us to establish a random guess baseline as a random coin flip since the sensitive attribute ``Hispanic'' is binary.

We focus on the scenario where the number of released aggregates, $m$, is smaller than the number of users $s$ in the protected dataset $D$, $m<s$ and randomly sample $m$ aggregate statistics without replacement from Table~\ref{tab:census_blocks}. 

The majority of the blocks are small, with $99\%$ of them containing less than $454$ records. We assume that the protected dataset $D$ is in the top 1 percent largest datasets. In order to create $D_{aux}$ and $D$ easily, we select the $10$ smallest blocks larger than $4540$ users. We partition them randomly into two parts of sizes 10\% and 90\%, representing the private dataset $D$ and the auxiliary dataset $D_{aux}$, respectively.

\textbf{American Community Survey} We conduct secondary experiments on the American Community Survey (ACS) dataset \cite{ACS}\footnote{In line with Dick et al.~\cite{dick2023confidence}, we use the folktable Python package \cite{ding2022retiringadultnewdatasets} to access the 1-year ACS dataset.}. The ACS dataset is partitioned by Public Use Microdata Areas (PUMAs), with a PUMA being the smallest region to contain record-level data. We choose similar non-sensitive attributes of the ACS dataset as the ones of the PPMF dataset, in particular PUMA-id, age, sex, and race. We take the attribute describing whether the income level of a person is over $\$50000$ as the sensitive attribute. Analogously to PPMF, we take the three smallest PUMAs with more than 4540 users, we split each dataset $90-10$ and randomize the values of the sensitive attribute.

In line with prior work~\cite{dick2023confidence}, we consider $k-way$ marginals as aggregate statistics on the ACS dataset. In particular, we consider all one-way and two-way marginals, such that we bucketize the age attribute into 5-year intervals.

\subsection{Evaluation Metrics}

\textbf{Area Under Receiver Operating Curve (AUC)}. We use the area under the receiver operating curve, denoted as AUC, to measure the strength of the attacks across all target users.

\textbf{True positive rate (TPR) at low false positive rate (FPR)}. As privacy is not an average-case metric\cite{carlini2022membership}, we measure--in line with the literature--the attacks' success to reliably infer the sensitive value of the most vulnerable users using True Positive Rate (TPR) at $k$ False Positive Rate (FPR), where $k$ is small. We denote this metric as $\text{TPR}@k\text{FPR}$.

\subsection{Attack Parameters}

\textbf{Parameters specific to state-of-the-art methods}. For both CIP and RAP, we obtain $K=100$ tentative reconstructed datasets, $D_1', \ldots, D_K'$, each of size $s$, as recommended by the authors of RAP~\cite{dick2023confidence}. For RAP, we generate each of the tentative reconstructed datasets with their proposed synthetic data generator model with an internal dimension of $1000$ and we allow each generator model to train over $1000$ iterations.

\textbf{Parameters specific to DeSIA}. We instantiate the number of shadow datasets in the stochastic module $\mathcal{M}_S$ to be $N=20 000$, $\frac{2}{3}$ of them are used as training shadow datasets, and $\frac{1}{3}$ as validation shadow datasets. We build the meta-classifier with logistic regression under L2 penalty. To reduce overfitting, we apply cross-validation combined with grid search to choose the best L2 penalty coefficient for the logistic regression. We allow the meta-classifier to train for maximally $1000$ iterations.

\textbf{General parameters}. If not otherwise specified, we sample $m$ aggregate statistics such that $\frac{m}{s}=0.25$.

%% file: sections/5results.tex
\section{Results}\label{sec:results}
\subsection{Attack Performance}

Figure~\ref{fig:tpr_fpr} (main) shows DeSIA to outperform all reconstruction-based methods at the attribute inference task, reaching an AUC of $0.75$. RAP-init is the second-best method with an AUC of $0.64$. Interestingly, while initialization using the auxiliary dataset $D_{aux}$ helps the RAP method, lifting the AUC from $0.59$ to $0.64$, it has no noticeable impact on CIP; both CIP-rand and CIP-init have an AUC of $0.63$. 

In line with the literature~\cite{carlini2022membership}, we also compare the performances of the methods focusing on the most vulnerable users. Figure~\ref{fig:tpr_fpr} (inset) shows DeSIA to strongly outperform all reconstruction-based methods here too, achieving a $\text{TPR}@10^{-3}\text{FPR}$ of $0.14$ while all other methods struggle to reliably predict the sensitive attribute of the most vulnerable users. 

\begin{figure}[h]
    \centering
    \includegraphics[width=1\linewidth]{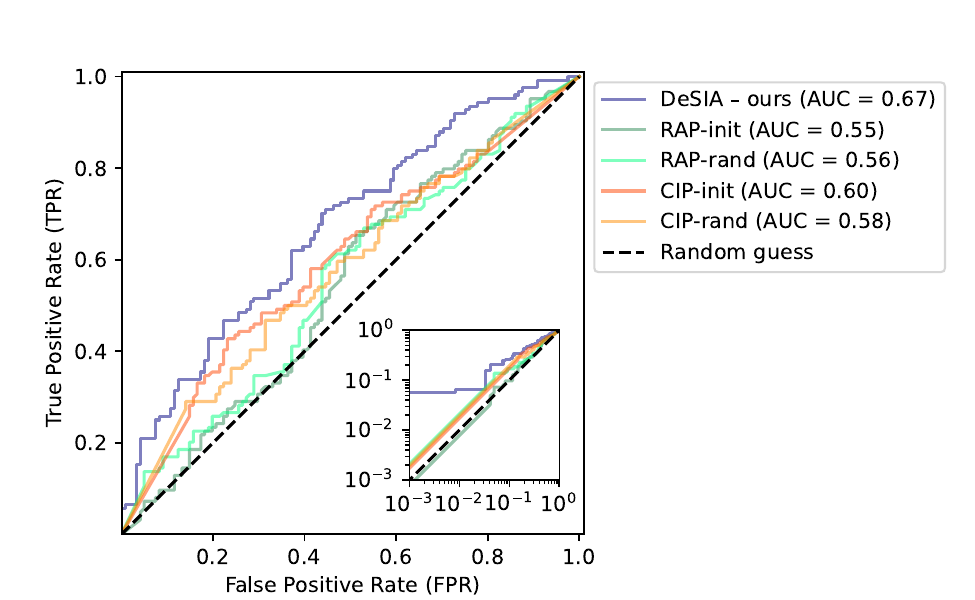}
    \caption{Performance on the PPMF dataset of our attack and the baseline reconstruction-based attacks at predicting the value of the sensitive attribute for only the target users who were not found to be deterministically vulnerable. The inset plot shares the same axes as the main plot.}
    \label{fig:tpr_fpr_non_vulnerable}
\end{figure}

\subsection{Attack Performance on Non-Deterministically Vulnerable Users}

In DeSIA, the deterministic module $\mathcal{M}_D$ identifies deterministically-vulnerable users, whose value of sensitive attribute $\mathcal{M}_D$ deterministically verifies to be the only possible value given the released aggregate statistics. These users are likely one of the easiest users to attack for all methods. We now compare the performance of all methods when excluding these target users. 

Figure~\ref{fig:tpr_fpr_non_vulnerable} (main) shows DeSIA to again outperform the reconstruction\hyp{}based methods, reaching an AUC of $0.67$ and maintaining a strong $\text{TPR}@k\text{FPR}$ even for those more difficult target users ($\text{TPR}@10^{-3}\text{FPR}$ of $0.06$). Interestingly, RAP suffers more than CIP from this exclusion with initialization having a marginal impact on the overall performance of both CIP and RAP.

\begin{figure}[t]
    \centering
    \includegraphics[width=1\linewidth]{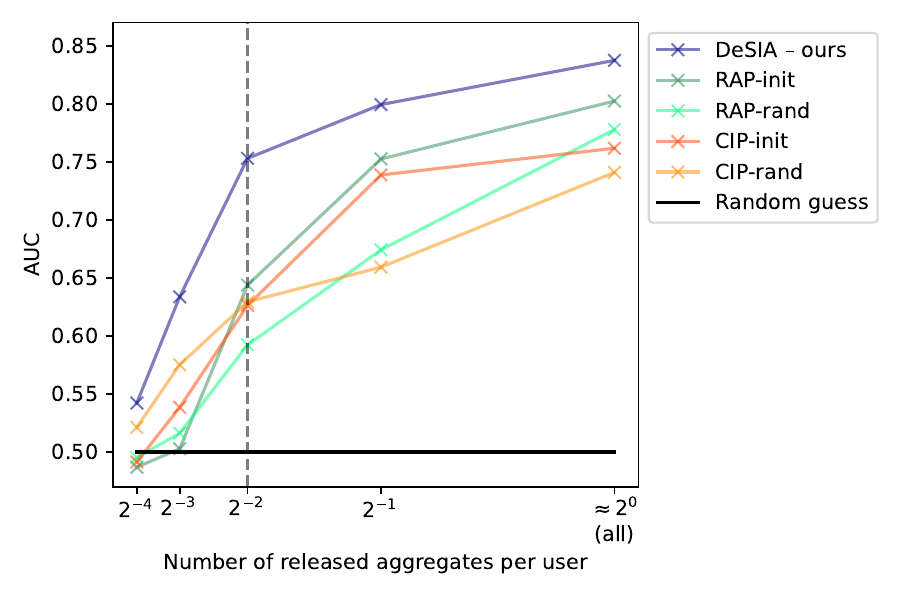}
    \caption{Impact of the number of released queries on the performance on the PPMF dataset of our attack and the baseline reconstruction-based attacks. The vertical dotted line indicates the default value taken for the other experiments.}
    \label{fig:ablation_num_queries}
\end{figure}

\begin{table*}[t]
    \centering
    \begin{tabular}{llll|cccc}
    \toprule
        \multicolumn{4}{c|}{\textbf{Method}} &  \textbf{AUC} & \multicolumn{2}{c}{\textbf{TPR@kFPR}} \\ 
        \midrule
        
        \textbf{Approach to find } &  \textbf{Uniqueness} & \textbf{Verification} & \textbf{Stochastic} & &  &  &  \\ 
        \textbf{a feasible value} &  \textbf{constraint} & \textbf{of feasible value} & \textbf{module} & & \textbf{k=10\%} & \textbf{k=1\%}   \\ \hline 
        solver & yes & yes & yes & 0.75  & 0.38 & 0.15  \\ \hline 
        solver & yes & yes & no & 0.63 &  0.09 & 0.09  \\ \hline 
        solver & yes & no & no & 0.59  & 0.0 & 0.0   \\ \hline 
        synthetic data & yes & no & no & 0.56   & 0.0 & 0.0   \\ \hline  
        synthetic data & no & no & no & 0.47 & 0.0 & 0.0  \\
    \bottomrule
    \end{tabular}
    \caption{Ablation study of the impact of different elements of our attack. The evaluation is performed on the PPMF dataset.}
    \label{tab:ablation}
\end{table*}

\subsection{Impact of the Number of Released Aggregates}\label{sec:num_queries}
We have so far evaluated the methods assuming that there are four times more users than the aggregate statistics in the protected dataset ($\frac{m}{s} = 0.25$). 
We here test the performance of DeSIA and reconstruction-based methods when different numbers of aggregate statistics are released. We sample different numbers of aggregate statistics at uniformly random from the same set of block-level contingency tables (see Table~\ref{tab:census_blocks}) with different ratios of $\frac{m}{s}$. We evaluate the aggregate statistics on the same protected dataset $D$.

Figure~\ref{fig:ablation_num_queries} shows DeSIA to outperform all reconstruction-based methods across all numbers of aggregate statistics. As expected, increasing the number of released aggregates increases the risk on average and lifts the performance of all methods. The state-of-the-art methods perform similarly among all sizes of aggregate statistics, with RAP being slightly better than CIP in general. 

\subsection{Ablation}\label{sec:ablation}
To better understand the contribution of each element of DeSIA, we perform an ablation study by systematically removing individual elements one by one and evaluating the impact on the method's performance. DeSIA has four main elements, three in the deterministic module $\mathcal{M}_D$ and one in the stochastic module $\mathcal{M}_S$. The three main elements of the deterministic module are: (1) finding a feasible value using either a solver or synthetic data generated by RAP, (2) adding the uniqueness constraint $c_t$ specific to the target user, and (3) verifying the uniqueness of the found feasible value. The inclusion of the stochastic module $\mathcal{M}_S$ is the fourth element. 

Table~\ref{tab:ablation} shows that all four elements influence DeSIA's performance and are necessary for the method to perform well. In particular, the removal of the stochastic module, replacing it with a random guess, and the removal of the verification of the feasible value substantially hurt the performance on AUC and TPR@kFPR. Our results emphasize the reliance of DeSIA on both the deterministic and the stochastic modules.

\subsection{Attack Performance on the ACS Dataset}
We have so far empirically evaluated DeSIA and state-of-the-art methods on the PPMF dataset. Here, we show their performance on another dataset released by the U.S. Census Bureau, the ACS dataset. In line with prior work~\cite{dick2023confidence}, we evaluate attacks on k-way marginal statistics. These are known to be a harder problem as the queries are less informative on average than the aggregate statistics from contingency tables carefully chosen by the U.S. Census Bureau (Table~\ref{tab:census_blocks}) on the PPMF dataset. 

\begin{figure}
    \centering
    \includegraphics[width=1\linewidth]{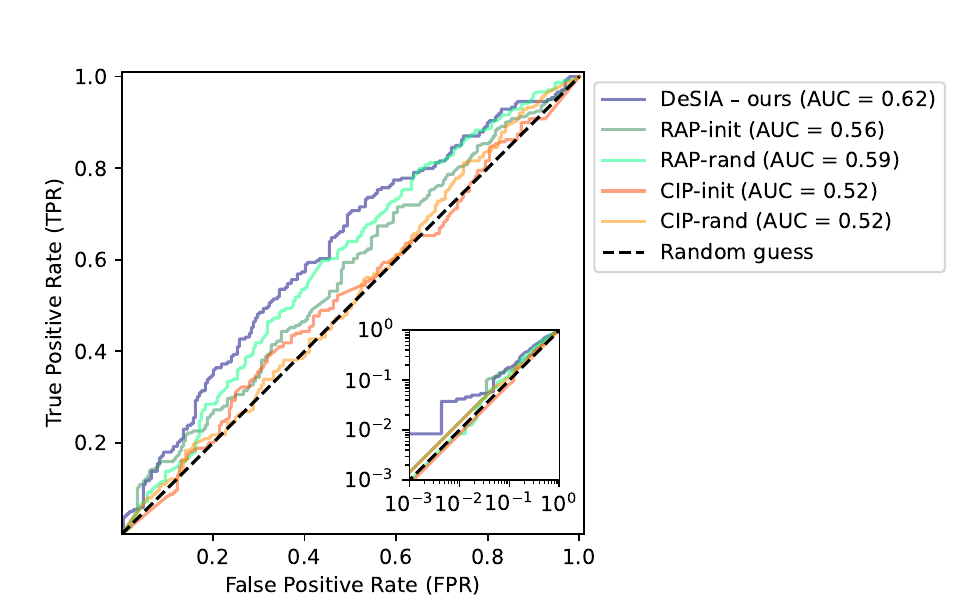}
    \caption{Performance on the ACS dataset of our attack and the baseline reconstruction-based attacks at predicting the values of sensitive attributes for all target users. The inset plot shares the same axes as the main plot.}
    \label{fig:tpr_fpr_ACS_dataset}
\end{figure}

Figure~\ref{fig:tpr_fpr_ACS_dataset} shows DeSIA to outperform all other methods both on overall AUC (main) and on TPR at low FPR (inset). Because of the less informative aggregate statistics compared to those on the PPMF dataset, the performance of all methods on ACS is lower than the performance on PPMF. 

\subsection{Robustness to Laplace Noise Addition}
In line with the literature~\cite{dick2023confidence,abowd20232010}, we have assumed that exact values of the aggregate statistics are released. Here, we perturb the released aggregate statistics by adding different levels of Laplace noise using a simple per-aggregate privacy budget. Given a per-aggregate budget of $\varepsilon$, we perturb the aggregates by adding Laplace noise $n_i\sim Laplace(\frac{1}{\varepsilon})$ to each aggregate statistic $q_i$. For simplicity and in line with U.S. Census Bureau's approach\cite{uscensusdas}, we round the answer to the nearest integer $round(q_i(D) + n_i), \forall i\in\{1,\ldots,m\}$. 

\begin{figure}[h]
    \centering
    \includegraphics[width=1\linewidth]{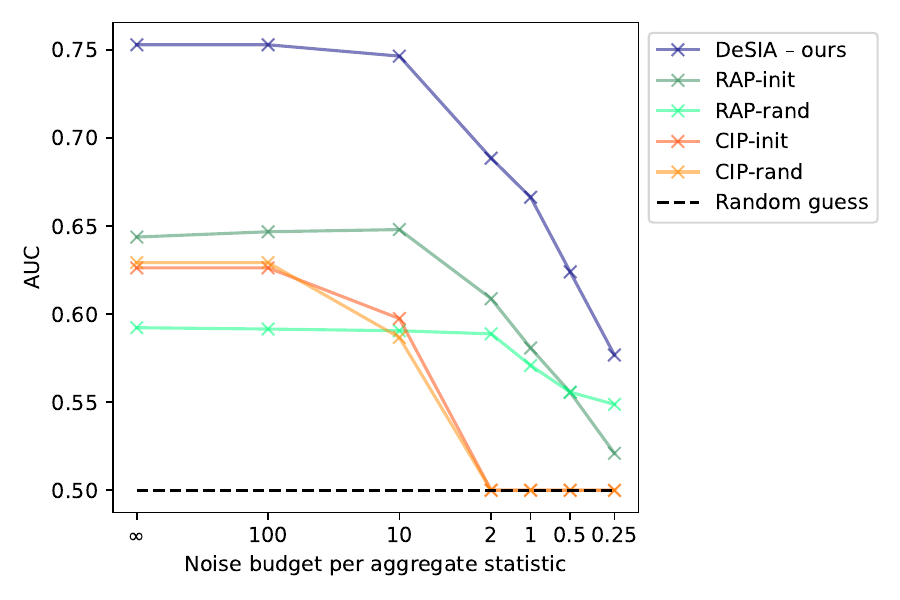}
    \caption{Robustness of the attacks to noise added to released aggregates. We report the mean AUC across $3$ independent noisy aggregate statistic releases. The evaluation is performed on the PPMF dataset.}
    \label{fig:noise}
\end{figure}

Figure~\ref{fig:noise} shows DeSIA to outperform all other methods and shows both DeSIA and RAP to be robust to even a significant level of noise addition; albeit at the cost of an expected drop in the performance of both methods. While CIP can handle some level of noise, it struggles to perform better than the random guess baseline when faced with larger levels of noise.  

%% file: sections/6MIA.tex
\section{Extension to Membership Inference Attack}

We have so far instantiated DeSIA on attribute inference attacks (AIAs), as we believe they to be a particular concern for aggregate data release. Membership inference attacks (MIAs) have, however, been extensively used to measure information leakage in the machine learning and synthetic data context and, in some cases, can be a concern in practice~\cite{grindr}. 

We here extend DeSIA to membership inference attacks (MIAs). We describe how we modify the privacy game to MIA, then extend DeSIA to MIAs, and compare it to reconstruct-based methods.

\subsection{Framework for Membership Inference Attack as a Privacy Game}

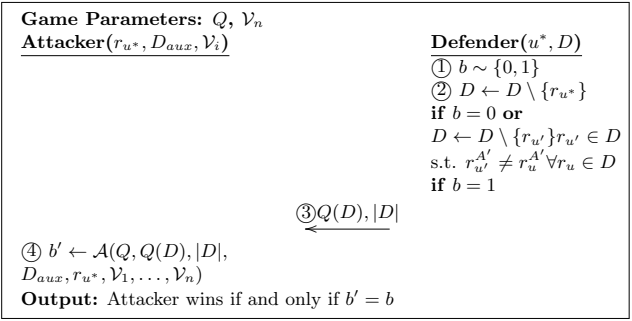
\begin{figure}[t]
\resizebox{.475\textwidth}{!}{
\fbox{\small
\begin{minipage}{1.2\columnwidth}
\begin{tabular}{lcl}
\textbf{Game Parameters: $Q$, $\mathcal{V}_n$} & & \\
\textbf{\underline{Attacker($r_{u^*}, D_{aux}, \mathcal{V}_i$)}}& &\textbf{\underline{Defender($u^*, D$)}} \\
 & & $\textcircled{1}$ $b \sim \{0, 1\}$ \\
 & & $\textcircled{2}$ $D \leftarrow D \setminus \{r_{u^*}\}$ \\
 & &  \textbf{if} $b=0$ \textbf{or} \\
& & $D \leftarrow D \setminus \{r_{u'}\} r_{u'} \in D$ \\
  & & s.t. $r_{u'}^{A'} \neq r_{u}^{A'} \forall r_u\in D$ \\ 
 & &  \textbf{if} $b=1$ \\
 & $\xymatrix@1@=40pt{& \ar[l]_*{\textcircled{3} Q(D), |D|}}$ & \\
 $\textcircled{4} \ b' \leftarrow \mathcal{A}(Q, Q(D),|D|,$ & & \\
 $D_{aux}, r_{u^*}, \mathcal{V}_1, \ldots, \mathcal{V}_n)$ & & \\
\end{tabular}
\begin{tabular}{l}
\textbf{Output:} Attacker wins if and only if $b' = b$
\end{tabular}
\end{minipage}
}}
\caption{Privacy game for membership inference attacks against fixed aggregate statistics.}
\label{fig:privacy_game_mia}
\end{figure}

We here modify the attribute inference attack privacy game to the membership inference case, randomizing the membership of the target users in the protected dataset instead of the sensitive attribute values. The game is as follows (Figure~\ref{fig:privacy_game_mia}).

First, the defender samples a secret bit $b$ uniformly at random from $\{0, 1\}$ to determine the membership of the target user $u^*$.

Second, the defender removes the target user from the private dataset $D$, if $b=0$, or removes another target user $u'$ from $D$, if $b=1$.

Third, the defender obtains the fixed aggregate statistics $Q(D)$ by evaluating the aggregate statistics $Q$ on the protected dataset $D$. The defender sends the values $Q(D)$ and the size of the protected dataset $|D|$ to the attacker.

Fourth, the attacker aims to infer the membership of the target user. The attacker runs an attack $\mathcal{A}$ and obtains a membership prediction $b'$.

The attacker wins the game if and only if the predicted membership bit $b'$ is equal to the secret bit $b$ held by the defender, $b' = b$.

\subsection{Extension of DeSIA to MIA}

We adapt DeSIA to MIA with small modifications on both the deterministic $\mathcal{M}_D$ and stochastic $\mathcal{M}_S$ modules. Note that, for the deterministic module $\mathcal{M}_D$, the uniqueness constraint $c_t$ is not applicable anymore, since the goal of $\mathcal{M}_D$ is to decide the deterministic membership. Furthermore, we verify the membership of the target user by finding a feasible multiplicity of target record $(x_{(r_{u^*}^1, \dots, r_{u^*}^{n})}$ and modifying the null constraint $c_t’$ to verify it: 
$$c_{t'} : x_{(r_{u^*}^1, \dots, r_{u^*}^{n})} = 1 - \sigma(x_{(r_{u^*}^1, \dots, r_{u^*}^{n})}).$$

For the stochastic module $\mathcal{M}_S$, we slightly change the generation of the shadow datasets to resemble the first two steps of the privacy game. Before creating each shadow dataset $D_{shadow}^i$, we now sample a variable $b^i$ uniformly at random from $\{0, 1\}$. If $b^i = 1$, we consider the target user to be a member of $D_{shadow}^i$ and sample $s-1$ users from the auxiliary dataset $D_{aux}$. If $b^i = 0$, we consider the target record to not be a member of $D_{shadow}^i$ and sample $s$ users from $D_{aux}$. We train the machine learning model $M$ on the aggregate statistics from these shadow datasets $\{(Q(D^1_{shadow}, b^1),  \ldots, (Q(D^N_{shadow}, b^N)\}$ to predict the membership of target user $\hat{b^i}$ given the input $Q(D)$.

\subsection{Adapting reconstruction-based attack to the MIA task}
Both reconstruction methods provide us with a number of tentative reconstructed datasets $D_1', \ldots, D_K'$. We adapt the AIA procedure (Section~\ref{sec:ra_to_aia}) to now perform a majority vote only from the 0-neighborhood of the target record $r_{u^*}$ across all tentative reconstructed datasets. If the target user appears in more than $\frac{K}{2}$ tentative reconstructed datasets, we consider $\hat{b} = 1$ and $\hat{b} = 0$ otherwise.

\subsection{Results}

\begin{figure}
    \centering
    \includegraphics[width=1\linewidth]{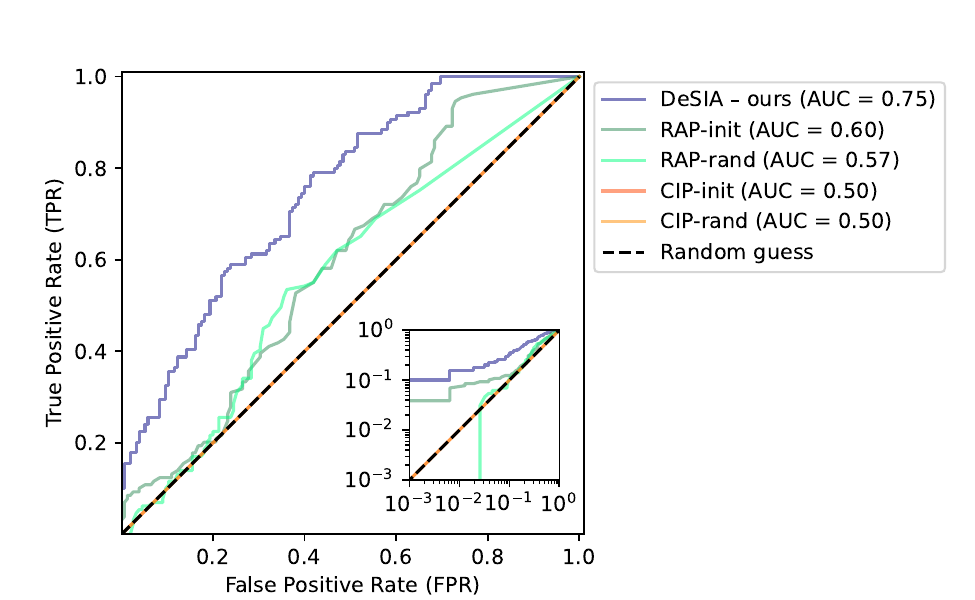}
    \caption{Membership inference attack performance on the PPMF dataset. The inset plot shares the same axes as the main plot. Interestingly, RAP-rand performs worse at TPR at low FPR on MIA than the random guess baseline. This is due to RAP incorrectly infers with high-confidence membership of users when their values for the non-sensitive attributes are relatively common on their own.}
    \label{fig:mia}
\end{figure}

Figure~\ref{fig:mia} shows DeSIA can be adapted to the MIA task, performing well both overall (AUC=0.85) and on the most vulnerable target record $\text{TPR}@10^{-3}\text{FPR}$ of $0.10$. While the RAP method can also be used for MIA it performs substantially worse than DeSIA while a straightforward adaptation of CIP does not allow it to perform better than a random guess.

%% file: sections/7discussion.tex
\section{Discussion}

\subsection{Using a Different Machine Learning Model}
To optimize for speed, the stochastic module $\mathcal{M}_S$ uses a logistic regression (LR) as a machine learning (ML) model to infer the sensitive attribute of the target user using aggregate statistics and its values. We here explore alternative options.

We evaluate three alternative choices of ML models:
\begin{itemize}
    \item \textbf{Multi-Layered Perceptron (MLP)}: an MLP with two hidden layers of sizes 50 and 20.

    \item \textbf{Random Forest (RF)}: a random forest model with $100$ trees. We set Gini impurity to be the criterion measuring the quality of the split. We select a random subset of features at each split, where the subset contains the square root of the total number of features.

    \item \textbf{Support Vector Machine (SVM)}: an SVM with radial-basis function kernels under hinge loss.
\end{itemize}

\begin{figure}
    \centering
    \includegraphics[width=1\linewidth]{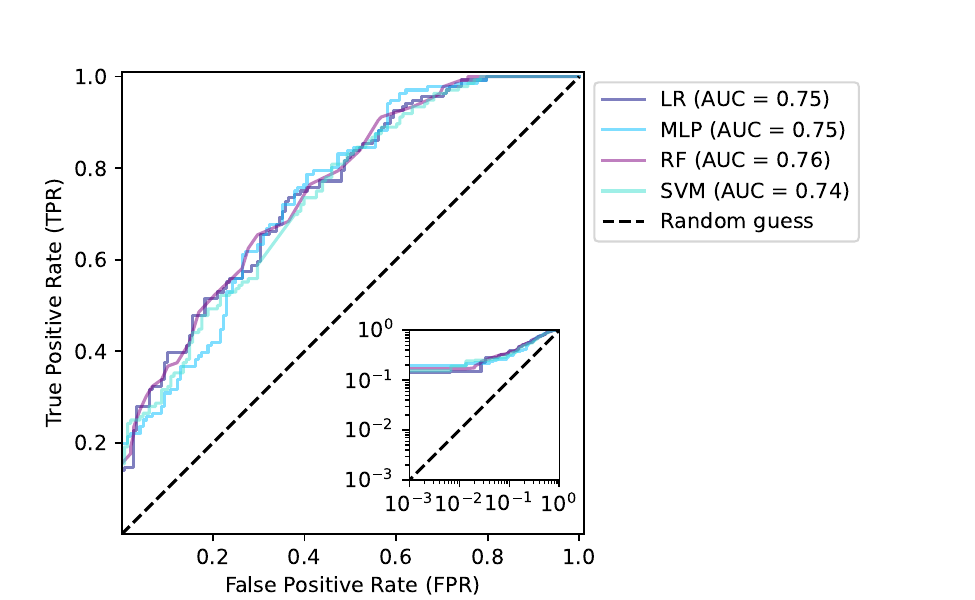}
    \caption{Performance on the PPMF dataset of our attack where the stochastic module $\mathcal{M}_S$ uses different machine learning models. The inset plot shares the same axes as the main plot.}
    \label{fig:diff_ml_models}
\end{figure}

Figure~\ref{fig:diff_ml_models} shows that the choice of the ML model for DeSIA's stochastic module does not have a substantial impact on the attack performance with all four ML models achieving an AUC of around $0.75$.

\subsection{Comparison to Likelihood Attack} 
We have thus far only considered using an ML model in the stochastic module $\mathcal{M}_S$. We here explore whether using an ML model is necessary or if it can be easily replaced with a simple likelihood attack.

We consider a simple likelihood attack to infer the value of the sensitive attribute for the target user by (a) inferring the most likely value of the sensitive attribute given each aggregate statistic independently and (b) performing a majority vote across aggregates to obtain the final inference.

To estimate the likelihood, we use the shadow datasets $D^1_{shadow}$, $\ldots, D^N_{shadow}$ sampled from the auxiliary dataset $D_{aux}$. We group the datasets into $|\mathcal{V}_n|$ groups, where each group $G_b$ contains shadow datasets with the same sensitive attribute value $b$ of the target user,
$G_b = \{D_{shadow}^i | z_{i,j}=b\}.$  For an aggregate statistic $q$, we calculate the mean $\mu_{q, b}$ and standard deviation $\sigma_{q, b}$ of the values on the shadow datasets in each group $G_b$, $\{q(D_{shadow}^i) z_{i,j}=b\}$. Given a the value $q(D)$ on the private dataset $D$, we estimate the likelihood of the sensitive value of the target user by assuming a Gaussian distribution of query answers and calculating: $$\text{Pr}[r_{u^*}^n = b] = \Phi(\frac{q(D) - \mu_{G_b}}{\sigma_{G_b}}),$$ where $\Phi$ is the probability density function of the Standard Normal Distribution. 

For an aggregate statistic $q$, we infer $r_{u^*}^n$ as the value $b$ from the group with the maximum likelihood. We obtain independent votes from all aggregates $q\in Q$ on which the membership of the target user $u^*$ in their userset $U_q$, $u^*\in U_q$, is dependent only on the value of the target user's sensitive attribute $r_u^n$, i.e., aggregates that have a condition on the sensitive attribute and do not exclude the target user's values in the conditions of the non-sensitive attributes, $Q_{u^*} = \{q \ |  r_{u^*}^i \in V^{q}_{i}, \ \forall i \in \{1, \ldots, n-1\} \text{ and } V_n^q \neq \mathcal{V}_n \}$. We perform a majority vote over the predictions of all aggregates in $Q_{u^*}$ to get the final prediction of $r_{u^*}^n$.

\begin{figure}[t]
    \centering
    \includegraphics[width=1\linewidth]{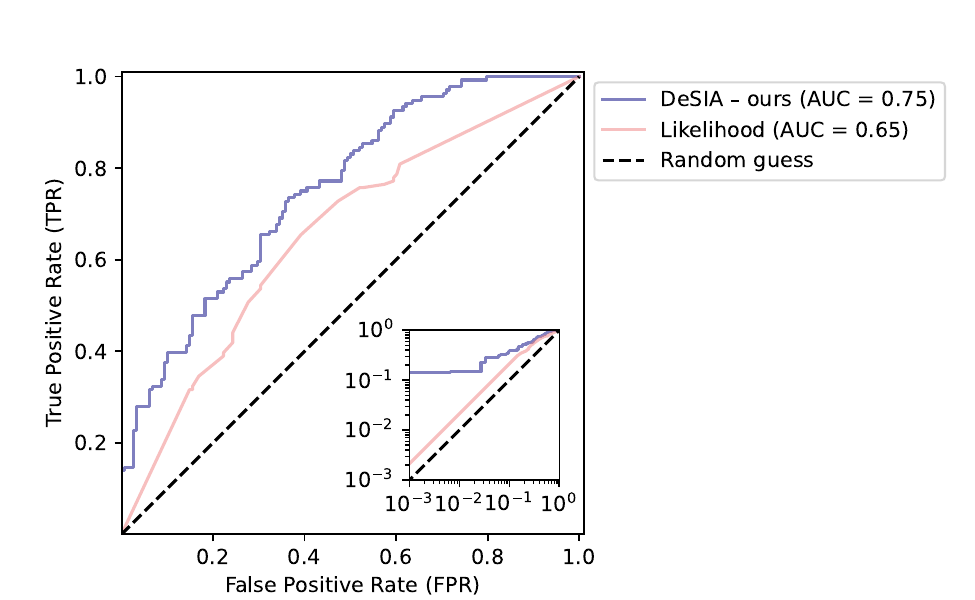}
    \caption{Performance on the PPMF dataset of our attack where the stochastic module $\mathcal{M}_S$ uses our standard machine learning model (Logistic Regression) and a likelihood attack. The inset plot shares the same axes as the main plot.}
    \label{fig:likelihood}
\end{figure}

Figure~\ref{fig:likelihood} shows that using an ML model is necessary and leads to a significantly better performing attack than the likelihood attack on both overall (AUC - main) and on highly vulnerable users (TPR$@k$FPR - inset). 

%% file: sections/8related_work.tex
\section{Related Work}\label{sec:related_work}
Empirical attacks have been proposed against a range of non-interactive data release mechanisms: fixed aggregate statistics, machine learning models, and synthetic data, as well as interactive release mechanisms, such as query-based systems.

\subsection{Attacks Against Fixed Aggregates}

\subsubsection{Inference Attacks}
The literature on inference attacks against aggregate statistics predates shadow model-based methods and often uses strong and theoretical assumptions. 

Homer et al.~\cite{homer2008resolving} introduced the first membership inference attack (MIA) against a large number of fixed aggregate statistics from genomic data. The attack has been later improved~\cite{jacobs2009new} and extended to microRNA data~\cite{backes2016membership}. Wang et al.~\cite{wang2009learning} extended the attack to an attribute inference attack (AIA). Finally,
Kasiviswanathan et al. \cite{kasiviswanathan2013power,kasiviswanathan2010price} introduced an attack against aggregate statistics from tabular data that relies on the so-called row-naming problem. It assumes that the users in the protected tabular dataset have unique identifiers that are both (a) known to the attacker and (b) used in the released aggregate statistics, which is not often the case in practice.

\subsubsection{Linear Reconstruction Attacks} Dinur et al. \cite{dinur2003revealing} proposed the first linear reconstruction attack which aims to reconstruct the protected dataset by solving a linear system of equations. 
Dwork et al. \cite{dwork2007price} extended their attack by relaxing the assumption that the noise added to the aggregates is bounded and introduced linear programming with error correction codes to reconstruction attacks. Following works improved the linear reconstruction attack by decreasing the requirement on the number of released aggregates \cite{dwork2008new} and improved the reconstruction of categorical attributes \cite{choromanski2012power}.

\subsubsection{State-of-the-art Reconstruction Attacks}
The U.S. Census Bureau introduced a constraint programming-based reconstruction attack \cite{abowd20232010} and instantiated it on the 2010 Decennial Census release. Dick et al. proposed a synthetic data-based reconstruction attack \cite{dick2023confidence} by solving a non-convex optimization problem~\cite{liu2021iterative}. We compare our attack to these two state-of-the-art attacks.

\subsection{Attacks Against Machine Learning Models}
An active field of research exists on empirical inference attacks against machine learning models. Attacks typically rely on the loss of the target model~\cite{yeom2018privacy} and shadow models~\cite{shokri2017membership,salem2018ml,choquette2021label}.

The first inference attacks against machine learning models were attribute inference attacks (also referred to in the literature as model inversion attacks~\cite{salem2018ml}). Fredrikson et al.~\cite{fredrikson2014privacy} proposed the first methodology and AIA against an ML model, inferring a patient's genetic marker from a model trained to predict the dosage of a drug. The attack was later extended from a black-box to a white-box attack by Mehnaz et al.~\cite{mehnaz2022your}. 
The correlation between the genetic marker and the dosage of the drug has, however, raised concerns about the validity of some of the results and the extent to which it proved that an information leakage was happening~\cite{githubBlogposts20160614mdMaster}. 
Jayaraman et al.~\cite{jayaraman2022attribute} formalized the issue and showed that popular black-box attacks often do not perform better than a baseline that uses these correlations to impute the values of the sensitive attribute. We refer to this as the imputation issue. Recent work on AIA against ML models, often image models, addressed this issue by evaluating the attack performance inferring the sensitive attributes with the model and without the model ~\cite{zhao2021feasibility}. In line with AIA against tabular datasets protected by query-based system (below) we address this issue by uniformly randomizing the sensitive attributes.

Soon after the first attribute inference attack was proposed, Shokri et al. proposed the first membership inference attack against machine learning models~\cite{shokri2017membership} and coined the term shadow models. This has become an active field of research with attacks such as RMIA~\cite{zarifzadeh2023low} and LiRA~\cite{carlini2022membership} and a focus on evaluating the performance of inference attacks on the most vulnerable users by reporting TPR at low FPR~\cite{carlini2022membership} which we follow.

Empirical attacks are often used to argue that models are not privacy-preserving by default, emphasizing the need for privacy-preserving measures, and to estimate the privacy risk in practice by models trained with DP privacy guarantees, typically using DP-SGD~\cite{abadi2016deep}. DP-SGD theoretical guarantees are indeed typically not tight, often overestimating the strength of the attacker~\cite{ponomareva2023dp,nasr2021adversary,stock2022defending}. The common approach when releasing machine learning models is thus to combine DP guarantees~\cite{ponomareva2023dp} with empirical attacks.

\subsection{Attacks Against Synthetic Tabular Data}
Contrary to machine learning models, synthetic data generators lack a notion of a loss for a given user. A separate line of research on attacks against released synthetic data, and up to a certain point against synthetic data generators including recently diffusion models, has thus developed. Attacks against synthetic data typically rely on the same shadow modeling technique and use the distance aggregate statistics as a signal, e.g., for membership. Stadler et al.~\cite{groundhog} proposed the first MIA against synthetic data. Houssiau et al.~\cite{houssiau2022tapas} and then Meeus et al.~\cite{meeus2023achilles} successively extended the attack by expanding the set of aggregate statistics used. In parallel, Annamalai et al.~\cite{annamalai2024linear} introduced an AIA against synthetic data by using a linear reconstruction attack. Most recently attacks targeting directly diffusion synthetic data generation models were proposed~\cite{wu2025winning}.

\subsection{Attacks Against Interactive Systems}
Query-based systems (QBSes) are interactive interfaces that allow analysts to send queries and retrieve aggregate answers about a protected dataset. Two broad types of QBSes have been implemented in practice: interfaces, such as TableBuilder~\cite{o2008table}, a special-purpose QBS for national census data by the Australian Bureau of Statistics, and general-purpose QBSes, such as Diffix~\cite{francis2017diffix}. 

\subsubsection{Reconstruction Attacks} 
Cohen et al. \cite{cohen2018linear} and Joseph et al.~\cite{differentialprivacyReconstructionAttacks} both proposed reconstruction attacks against Diffix~\cite{francis2017diffix}. Both attacks were based on the earlier work of Dinur et al.~\cite{dinur2003revealing}.

\subsubsection{Membership Inference Attacks}
Pyrgelis et al.~\cite{benthamsgazeLocationTime} proposed an MIA against a query-based system protecting spatio-temporal data, based on their earlier work~\cite{pyrgelis2017knock}.

\subsubsection{Attribute Inference Attacks}
Chipperfield et al. \cite{chipperfield2016australian} and Rinott et al. \cite{rinott2018confidentiality} proposed AIAs against TableBuilder, exploiting the seeded bounded noise.
Gadotti et al.~\cite{gadotti2019signal} proposed an AIA exploiting Diffix's noise structure. Cretu et al.~\cite{cretu2022querysnout} proposed an automated approach to discovering AIAs against query-based systems showing it to outperform the best known manual attacks. Stevanoski et al.~\cite{stevanoski2024querycheetah} then further improved the speed of the attack and the space of sets of queries that can be searched.

%% file: sections/9conclusion.tex
\section{Conclusion}
We propose to a framework for inference attacks against fixed aggregate statistics. We introduce DeSIA, a deterministic stochastic attribute inference attack against fixed aggregate statistics. We instantiate the attack on fixed aggregates released by the U.S. Census Bureau.

We first show DeSIA to be highly effective at the attribute inference task and to strongly outperform reconstruction-based approaches. This is true both when predicting the values of the sensitive attribute overall (AUC), but also when predicting the values of the sensitive attribute for the most vulnerable users (TPR@kFPR).

Second, we show DeSIA to perform well when we vary (a) the numbers of released aggregates, (b) the dataset and aggregates, and (c) the levels of noise addition. 

Third, we perform an ablation study of the different components of our method, and show all components to be important to the overall performance. In particular, we show the stochastic module to be important for both the overall AUC and TPR at low FPR, and the verification by the deterministic module to be important for identifying highly-vulnerable users with only one possible value for their sensitive attribute. 

Fourth, we extend DeSIA to membership inference attack and show it to also significantly outperform all other methods at the MIA task both on overall AUC as well as on identifying highly-vulnerable users with high TPR at low FPR.

Our results show how even a limited number of aggregate statistics can be at risk from privacy attacks. Taken together, they strongly emphasize the need to implement formal privacy mechanisms and to empirically test the privacy of the data release in practice.

%% file: sections/acknowledgements.tex
\section*{Acknowledgments}
This work has been partially supported by the CHEDDAR: Communications Hub for Empowering Distributed ClouD Computing Applications and Research funded by the UK EPSRC under grant numbers EP/Y037421/1 and EP/X040518/1 and by the Agence E-Sant\'e  Luxembourg~\footnote{https://www.esante.lu/}. We acknowledge computational resources provided by the Imperial College Research Computing Service.~\footnote{http://doi.org/10.14469/hpc/2232}